\documentclass[oupdraft]{bio}
\usepackage{url}
\usepackage[utf8]{inputenc}
\usepackage{mathtools}
\usepackage{url}
\usepackage{amsmath}
\usepackage{amsfonts}
\usepackage{graphicx}
\usepackage{subfigure}
\usepackage{booktabs} 
\usepackage{rotating}
\usepackage{natbib}
\usepackage{thmtools}
\declaretheorem[style=definition,numberwithin=section]{procedure}

\begin{document}

\title{Faster estimation for constrained gamma mixture models using closed-form estimators}

\author{JIANGMEI XIONG, ELIOT MCKINLEY, JOSEPH ROLAND,\\
ROBERT COFFEY, MARTHA J.SHRUBSOLE, KEN. S. LAU,\\SIMON N.VANDEKAR\\[4pt]
\textit{2525 West End Avenue, Room 1136, Nashville, Tennessee, US}
\\[2pt]
{simon.vandekar@vumc.org}}

\markboth%
{J. Xiong and others}
{Closed-form Gamma Mixture Model}

\maketitle


\begin{abstract}
{Mixture models are useful in a wide array of applications to identify subpopulations in noisy overlapping distributions. For example, in multiplexed immunofluorescence (mIF), cell image intensities represent expression levels and the cell populations are a noisy mixture of expressed and unexpressed cells.  Among mixture models, the gamma mixture model has the strength of being flexible in fitting skewed strictly positive data that occur in many biological measurements. However, the current estimation method uses numerical optimization within the expectation maximization algorithm and is computationally expensive. This makes it infeasible to be applied across many large data sets, as is necessary in mIF data. Powered by a recently developed closed-form estimator for the gamma distribution, we propose a closed-form gamma mixture model that is not only more computationally efficient, but can also incorporate constraints from known biological information to the fitted distribution. We derive the closed-form estimators for the gamma mixture model and use simulations to demonstrate that our model produces comparable results with the current model with significantly less time, and is excellent in constrained model fitting.}
{EM algorithm; Bayesian Statistics; Clustering; Single-cell Imaging}
\end{abstract}

\section{Introduction}
\label{sec1}

Mixture models are ubiquitous tools across scientific fields that can be used for clustering observations, because they are accessible and interpretable for researchers without extensive background in statistics.
Variations of mixture models are widely used to identify patient subgroups or cell populations and to identify tissue types or tumors in medical imaging \citep{geuenich_automated_2021,zhang_segmentation_2001,ashburner_unified_2005,solomon_segmentation_2006}.
In multiplexed immunofluorescence (mIF) imaging data, up to 60 markers are stained on a single tissue and individual cells are segmented using machine learning \citep{mckinley_optimized_2017,gerdes_highly_2013}.
The histogram of cell marker intensities is an overlapping mixture of background intensity and cells that are expressed for the marker.
For example, the marker CD8 is expressed only in T cells that are cytotoxic T cells and the majority of cells express background levels of CD8. Mixture models can be used to identify which cells are positive for CD8.

Normal mixture models are most commonly used as they have closed-form solutions for multivariate data.
However, these models are not appropriate for mixtures of positive, heavily skewed distributions, which are common in biological imaging.
For example, mIF imaging data are a mixture of heavily skewed distributions, where the components closer to zero are cells that do not express the imaged marker. In these cases, a gamma mixture model (GMM) has better flexibility to fit these heavily skewed distributions \citep{young_finite_2019,benaglia_mixtools_2009,khalili_gamma-normal-gamma_2007}.

To obtain estimates for parameters, mixture models introduce the complete data loglikelihood and use the expectation maximization (EM) algorithm to iterate between taking the expectation of the unknown latent variables and maximizing the log-likelihood conditional on the expected values \citep{dempster_maximum_1977}.
A major drawback of GMMs is that they do not have closed-form maximum likelihood estimates (MLEs) and computation can take a long time \citep{benaglia_mixtools_2009,young_finite_2019,lakshmi_parameter_2016}.
The state-of-art approach is to use expectation conditional maximization to obtain estimates for the parameters \citep{young_finite_2019}.
This approach (as implemented in the {\tt R} package {\tt mixtools}) requires numerical root finding to obtain an estimate for the parameters in every maximization step of the EM algorithm.
This slow maximization makes the approach infeasible for large data sets, where the algorithm must be applied a large number of times, such as in mIF imaging, where the algorithm is applied to many channels on a large number of slides.
The lack of a closed-form estimate also makes it difficult to incorporate prior biological knowledge that may improve model fit.

Recently, \citet{ye_closed-form_2017} developed closed-form estimators for parameters for the gamma distribution by utilizing the generalized gamma distribution.
Here, we use this approach to derive closed-form estimators for the gamma mixture model (cfGMM) using the expectation-maximization algorithm.
Our approach significantly reduces the computation time due to the estimators being closed-formed, and easily incorporates inequality constraints on the mode of the mixture components, which enables the addition of prior biological information.
We compare the cfGMM and constrained cfGMM to the maximum likelihood (ML) approached used in the {\tt mixtools R} package.
We use two simulations to demonstrate the cfGMM has similar bias and variance to the GMM, faster convergence times, and more frequently converges successfully.
Because of these advantages, the cfGMM approach can replace the current GMM estimation in most applications.
We use the two component constrained cfGMM to fit CD8 marker distributions and identify the proportion of CD8+ cells.
We provide code to fit the cfGGM in an {\tt R} repository on github (\url{https://github.com/JiangmeiRubyXiong/cfGMM}).

\section{Methods}
\label{sec2}

\subsection{Closed-form estimators for finite gamma mixture models}\label{sec2.2}

The MLE for shape and scale parameter of the gamma distribution has no closed-form expression \citep{ye_closed-form_2017}.
The MOM estimator has substantially higher variance than the MLE so is not an appealing alternative.
As a convenient solution, Ye and Chen \citep{ye_closed-form_2017} derived closed-form estimators for the gamma parameters using a generalized gamma distribution \citep{stacy_generalization_1962}
\begin{equation}
f(x)=G(a,b,\gamma)=\frac{ x^{a\gamma-1}\exp\{(-x/b)^{\gamma}\}}{b^{a\gamma}\Gamma(a)}. \label{eq:generalizedGamma}
\end{equation}
When $\gamma=1$, \eqref{eq:generalizedGamma} is equal to the gamma distribution with shape $a$ and scale $b$. Closed-form estimators of $a$ and $b$ are obtained by differentiating \eqref{eq:generalizedGamma} with respect to $b$ and $\gamma$, setting the derivative to 0, solving for $a$ and $b$, and setting $\gamma=1$ in the final expression.

To derive the closed-form estimators for the GMM, we assume the data is a random sample $x_1, \ldots, x_n$ from a $K$ component generalized gamma mixture distribution. The density function of $X$ is
\begin{equation*}
P(X=x)=\sum_{k=1}^K \lambda_k f(x; a_k, b_k, \gamma_k).
\end{equation*}
and the log-likelihood of the dataset is
\begin{equation}
    \ell(\mathbf{x}|\mathbf{a},\mathbf{b},\pmb{\lambda})=\sum^n_{i=1}\log\left\{\sum^K_{k=1}\lambda_kf(x_i|a_k,b_k)\right\}
    \label{eq:loglikelihood}
\end{equation}
For each generalized gamma component $k$, $\lambda_k\in [0,1]$ are the mixture parameters, $\sum_{k} \lambda_k = 1$; $f$ denotes the generalized gamma density function; $a_k, b_k, \gamma_k$ are the parameters for the generalized gamma \eqref{eq:generalizedGamma}.

The expectation maximization (EM) algorithm \citep{dempster_maximum_1977} is a standard approach for parameter estimation in mixture models.
This approach introduces the latent multinomial variable $Z_{i} = (Z_{i1}, \ldots, Z_{iK})$ into the model and maximizes the expected value of the complete data likelihood \citep{dempster_maximum_1977}.
The expectation of the complete data likelihood to be maximized for the generalized gamma distribution is
\begin{equation*}
\mathbb{E}_Z \ell(x \mid Z) = \sum_{i=1}^n \sum_{k=1}^K z_{ik} \log f(x_i; a_k, b_k,\gamma_k),
\end{equation*}
 where 
 \begin{equation}
 \label{eq:lambdaFormula}
    z_{ik} = \mathbb{P}(Z_{ik}=1 \mid x_i;\pmb{ a, b, \gamma} ) =\frac{f(x_i|a_k, b_k, \gamma_k)}{\displaystyle\sum^K_{j=1}f(x_i|a_j, b_j, \gamma_k)},
 \end{equation}
$\mathbf{a} = (a_1, a_2, \ldots, a_K)$, and $\mathbf{b}$, $\pmb \gamma$, and $\pmb \lambda$ in \eqref{eq:loglikelihood} are similarly defined vectors.

From here, the maximization of the expectation is now analogous to the maximization of generalized gamma distribution for each component of the mixture model.
Following the steps by \citet[][see Appendix]{ye_closed-form_2017}, note again that $\gamma$ is set to $1$ in the end, the expression of closed-form estimator for each component can be derived as

\begin{align}
    \hat {a}_k & =\frac{\displaystyle\sum^{n}_{i=1}z_{ik}\displaystyle\sum^{n}_{i=1}z_{ik}X_i}{\displaystyle\sum^{n}_{i=1}z_{ik}\displaystyle\sum^{n}_{i=1}z_{ik}X_i\log X_i-\displaystyle\sum^{n}_{i=1}z_{ik}\log X_i\displaystyle\sum^{n}_{i=1}z_{ik}X_i}\label{eq:ak} \\
    \hat {b}_k& =\frac{\displaystyle\sum^{n}_{i=1}z_{ik}\displaystyle\sum^{n}_{i=1}z_{ik}X_i\log X_i-\displaystyle\sum^{n}_{i=1}z_{ik}\log X_i\displaystyle\sum^{n}_{i=1}z_{ik}X_i}{\left(\displaystyle\sum^{n}_{i=1}z_{ik}\right)^2}\label{eq:bk}\\
    \hat {\lambda}_k & =\frac{\displaystyle\sum^{n}_{i=1}z_{ik}}{n} \label{eq:lambdak}
\end{align}
We can see from \eqref{eq:ak}, \eqref{eq:bk}, \eqref{eq:lambdak} that the estimators depend on $z_{ik}$, and $z_{ik}$ depend on the unknown parameters.
We proceed with the classical EM estimation approach as follows \citep{dempster_maximum_1977}.

\begin{procedure}[cfGMM EM algorithm] Let $K$ denote the number of components, $M$ denote the maximum number of interations, and $\epsilon>0$ be the convergence criterion.
\begin{enumerate}
        \item Initialize $C = \infty$, $t=1$. Randomly generate $\pmb{\lambda}^{(0)}$ so that $\sum_k \lambda_k=1$. Initialize parameter vectors of $\textbf{a}^{(0)}, \textbf{b}^{(0)}\in \mathbb{R}^K$ using method of moments (MOM) estimator of the data partitioned according to $\pmb{\lambda}^{(0)}$ \citep{young_finite_2019}. 
    
    \item While $t<M$ and $C>\epsilon$, compute the expected values $z_{ik}^{(t)}$ using formula \eqref{eq:lambdaFormula}, with $\textbf{a}=\textbf{a}^{(t-1)}$ and $\textbf{b}=\textbf{b}^{(t-1)}$.
    Then compute $\textbf{a}^{(t)}$, $\textbf{b}^{(t)}$ and $\pmb{\lambda}^{(t)}$ using formulas \eqref{eq:ak}, \eqref{eq:bk} and \eqref{eq:lambdak}  with $z_{ik} = z_{ik}^{(t)}$. Use formula \eqref{eq:loglikelihood} to set $C=\big|\ell(\mathbf{x}|\mathbf{   a}^{(t)},\mathbf{b}^{(t)},\pmb{\lambda}^{(t)})-\ell(\mathbf{x}|\mathbf{a}^{(t-1)},\mathbf{b}^{(t-1)},\pmb{\lambda}^{(t-1)})\big|/n$, $t= t+1$. Repeat until convergence.
    
    \item Set $\hat{\textbf{a}}= \textbf{a}^{(t)}, \hat{\textbf{b}} = \textbf{b}^{(t)}$, and $\hat{\pmb{\lambda}} =  \pmb{\lambda}^{(t)}$.
    
\end{enumerate}
\label{proc:cfGMM}
\end{procedure}

Because our closed-form estimators are more efficient than the classical approach, our algorithm has the option to restart with several random starting values and choosing the one with the largest log-likelihood value to avoid reporting estimates obtained at a local maximum. This option is strongly encouraged especially for model with more than 2 components due to model complexity \citep{karlis_choosing_2003}.
As in other software implementations \citep{benaglia_mixtools_2009}, if parameter values diverge (e.g. go to infinity) then the whole algorithm will restart with new starting values.
Details are available in our code on our github repository \url{https://github.com/JiangmeiRubyXiong/cfGMM}.

\subsection{Constrained Estimator}
In many applications, we have prior information that the mode of components are within an certain area.
For example, in mIF imaging data we know that, for the CD8 marker, there is a small proportion of CD8+ cells whose mode should be to the right of the unexpressed cells. 
To incorporate this information, we constrain the mode, $m_k$, which is a function of parameters $a_k,b_k$, to be in an interval
\begin{equation} \label{eq:mode_bound}
  m_k=(a_k-1)\frac{1}{ b_k }\in (l_k,u_k) , k=1,2,3,...,K
\end{equation}

This constraint is equivalent to putting a flat bounded prior on the mode of the components. Because the the log-likelihood is strictly concave with respect to $a_k$ and $b_k$, the constrained maximum must lie on the boundary, if the global maximum is outside the bounds \citep{boos_essential_2013}.
In this case, the expected log-likelihood of the component $k$ is maximized at $m_k$
\begin{equation}\label{eq:bound}
    (a_k-1)\frac{1}{ b_k } = m_k \Rightarrow a_k=m_k\frac{1}{ b_k }+1
\end{equation}

\begin{figure}[!p]
\centering\includegraphics{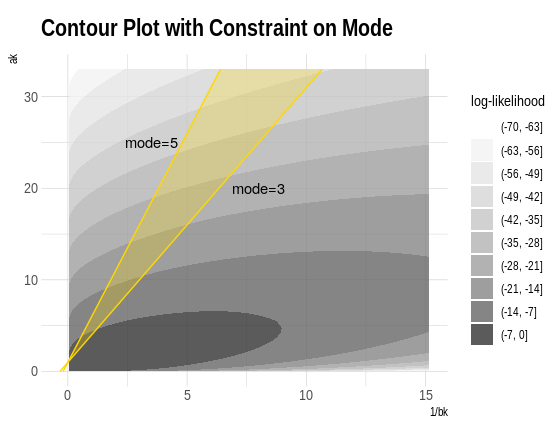}
\caption{A contour plot of log-likelihood with mode constraints. The log-likelihood is created using one of the two-component simulation data sets. The straight lines are the $m_k=3, m_k=5$ boundary, separately and their expression follows equation \eqref{eq:bound}. The range allowed for mode is the shaded area between the line of mode bounds. Due to concavity of the likelihood function the constrained maximum occurs on the boundary.}
\label{fig:likelihood}
\end{figure}

Figure \ref{fig:likelihood} is a graphical depiction of the boundaries and likelihood, where the likelihood is generated from one of 2-component datasets from the simulation section.  To find the constrained MLE of $b_k$, we can plug equation \eqref{eq:bound} in the expected log-likelihood with $\gamma_k=1$, take derivative with respect to $b_k$, and set it equal to zero
\begin{equation}\label{eq:unsolvedpartial}
    \frac{\partial \mathbb E_{z|x}[\log(L(\mathbf{x}|\mathbf{z}))]}{b_k}=
    \sum^{n}_{i=1}\frac{z_{ik}}{-b_k^2}\bigg(m_k+b_k-m_k\log(b_k)-m_k\psi(m_k/b_k+1)+m_k\log X_i - X_i.
    \bigg)=0
\end{equation}

Unfortunately, this equation does not have a closed-form solution due to the digamma function, $\psi(.)$. Therefore, we use Newton-Raphson to solve for the optimal $b_k$.
This approach gives the following procedure.

\begin{procedure}[constrained cfGMM EM algorithm] Let $K$ denote the number of components, $M$ denote the maximum number of interations, and $\epsilon>0$ be the convergence criterion.
\begin{enumerate}
    \item Initialize $C = \infty$, $t=1$. Randomly generate $\pmb{\lambda}^{(0)}$ so that $\sum_k \lambda_k=1$. Initialize parameter vectors of $\textbf{a}^{(0)}, \textbf{b}^{(0)}\in \mathbb{R}^K$ using method of moments (MOM) estimator of the data partitioned according to $\pmb{\lambda}^{(0)}$.

    \item While $t<M$ and $C>\epsilon$
        \begin{enumerate}
            \item Compute $z_{ik}^{(t)}$ using formula \eqref{eq:lambdaFormula}, with $\textbf{a}=\textbf{a}^{(t-1)}$ and $\textbf{b}=\textbf{b}^{(t-1)}$.
            \item Compute $\textbf{a}^{(t)}$, $\textbf{b}^{(t)}$ and $\pmb{\lambda}^{(t)}$ using formulas \eqref{eq:ak}, \eqref{eq:bk} and \eqref{eq:lambdak}  with $z_{ik} = z_{ik}^{(t)}$. Check if equation \eqref{eq:mode_bound} holds for all pairs of $a_k^{(t)}, b_k^{(t)}$. If not, set $m_k$ to its closest boundary, and plug it in equation \eqref{eq:unsolvedpartial} to solve for $b_k$, and set $a_k$ by equation \eqref{eq:bound}.
            \item Using formula \eqref{eq:loglikelihood}, set $C=\big|\ell(\mathbf{x}|\mathbf{a}^{(t)},\mathbf{b}^{(t)},\pmb{\lambda}^{(t)})-\ell(\mathbf{x}|\mathbf{a}^{(t-1)},\mathbf{b}^{(t-1)},\pmb{\lambda}^{(t-1)})\big|/n$ and $t= t+1$.
        \end{enumerate}
         Repeat until convergence.
    
    \item Set $\hat{\textbf{a}}= \textbf{a}^{(t)}, \hat{\textbf{b}} = \textbf{b}^{(t)}$, and $\hat{\pmb{\lambda}} =  \pmb{\lambda}^{(t)}$

\end{enumerate}
\label{proc:cstr.cfGMM}
\end{procedure}
Using the constrained estimator can not only incorporate external information, but is also helpful in avoiding the algorithm from being stuck in local maximum of log-likelihoods.
Using correctly specified constraints can also reduce computation time with comparable result output as the unconstrained one.  

\subsection{Simulation Setup}
\label{sec:simsetup}

\begin{figure}[!p]
\centering\includegraphics{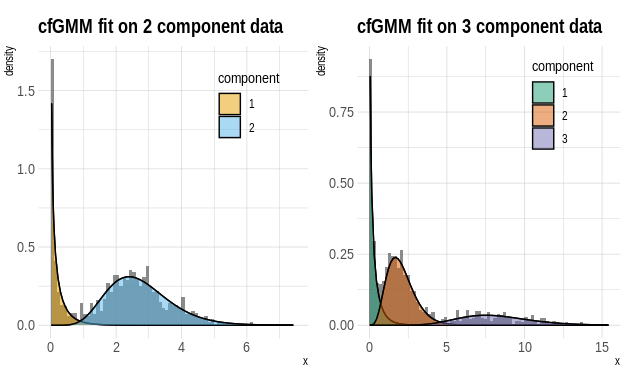}
\caption{Example cfGMM fit on one 2-component and 3-component simulation. Note that the histogram in the plot is the histogram of the data simulated according to the parameter setup (Section \eqref{sec:simsetup}), and the shaded curve is the density according to the cfGMM fit results.}
\label{fig:goodfit}
\end{figure}

We compare the cfGMM and the constrained cfGMM to the existing GMM, using two simulated datasets for 1,000 simulations each.
For the first data set we simulated a two component mixture model with parameters $\pmb \lambda = (0.3, 0.7)$, $\pmb \alpha = (0.5, 8)$, $\pmb \beta = (0.5, 1/3)$ and a three component mixture model with parameters $\pmb \lambda = (0.3, 0.5, 0.2)$, $\pmb \alpha = (0.5, 6, 8)$, $\pmb \beta = (2,1/3,1)$ (Figure \ref{fig:goodfit}).
For the constrained estimator, we restricted the mode of each component to be in the range $(-\infty, 0),(0,5)$, for the two component data, and $(-\infty, 0),(0,5),(5,15)$ for three component data. Note that they included the true mode for each component, which is $-\infty, 7/3$ for two component data and $-\infty, 5/3, 7$ for three component data. Note that there is no mode for gamma distribution with $\alpha<1$, and in our simulation the mode with $\alpha<1$ is set to be $-\infty$ for the convenience of calculation and representation.
We perform the simulation for sample sizes of 100, 1,000 and 10,000. We assessed bias, variance, computing time, and convergence rates to compare the methods.

\section{Simulation Results}
\label{sec3}

We simulated 2-component and 3-component gamma mixture models to compare convergence rate, convergence time, bias, and variance.
An example of the simulated data with the density of fitted model is given in Figure \ref{fig:goodfit}.

Our closed-from estimator has a higher proportion of convergence: all runs have converged in the simulation (Table \ref{tab:convergence}). In comparison, GMM is a lower convergence rate for the 3-component model.

\begin{figure}[!p]
\centering\includegraphics[height=3.9in]{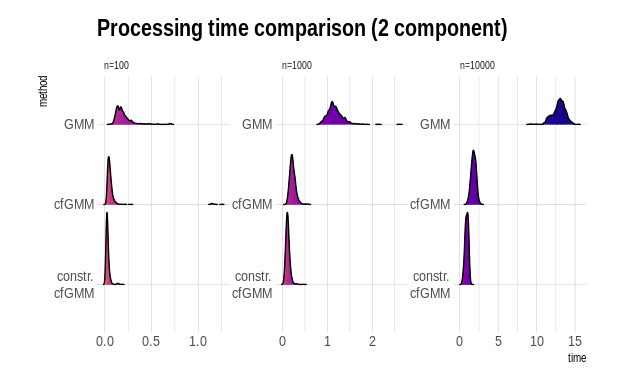}
\includegraphics[height=3.9in]{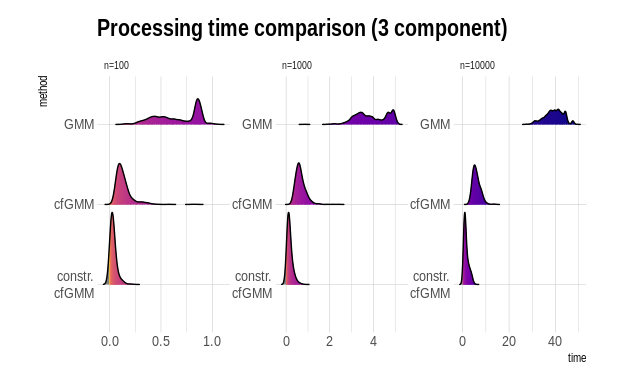}
\caption{A comparison of processing times for the two component (top) and three component (bottom) simulations. cfGMM and cfGMM with constraints both outperform GMM in computation time.}
\label{fig:processingTime}
\end{figure}

We compared convergence time across the 1000 simulations cfGMM outperforms GMM by great deal, and the constrained cfGMM is faster than non-constrained GMM (Figure \ref{fig:processingTime}). The difference between computation time increases as the sample size increases, so the cfGMM has a greater benefit in larger samples. 

\begin{figure}[!p]
\centering\includegraphics{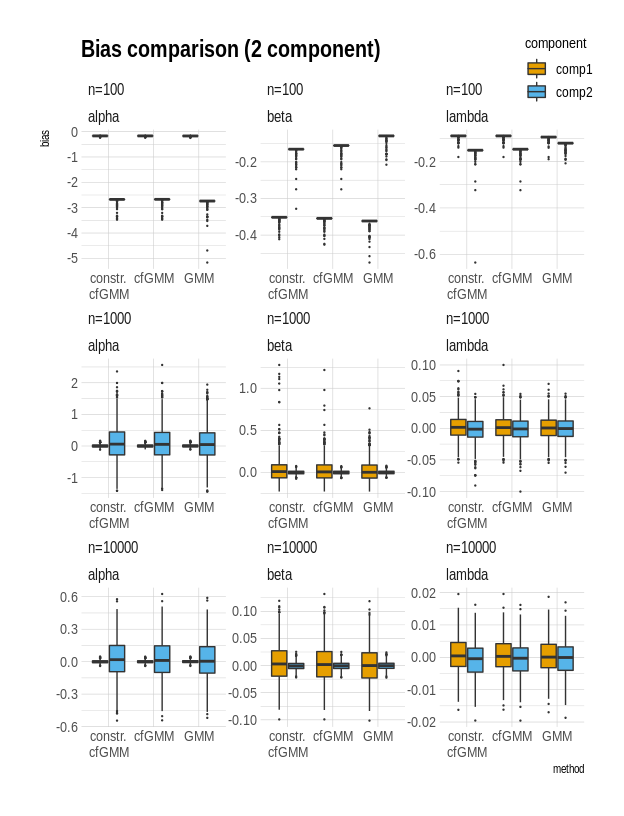}
\caption{Bias across 1,000 simulations of the two component model. The bias are winsorized (95\%) for display purposes. In small samples, all estimators are biased, but the bias is reduced in larger samples.}
\label{fig:bias2}
\end{figure}

\begin{figure}[!p]
\centering\includegraphics{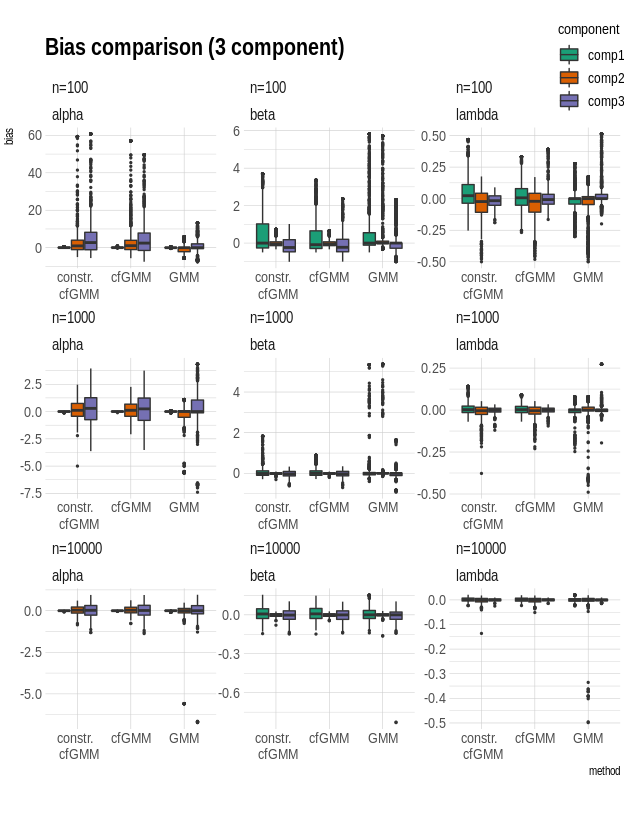}
\caption{Bias across 1,000 simulations of the three component model. The bias are winsorized (95\%) for display purposes. In small samples, all estimators are biased, but the bias is reduced in larger samples.}
\label{fig:bias3}
\end{figure}

 Both constrained and unconstrained cfGMM have bigger bias and variability in parameter estimation than GMM when $n=100$, but the discrepancy is diminished as the sample size increases (Figure \ref{fig:bias2} and \ref{fig:bias3}).  

\section{Identifying CD8$+$ cells mIF}
\label{sec4}

To compare our cfGMM with constraints to the classical GMM estimation method we fit the two models to CD8 expression data acquired using mIF imaging.
Prior to analysis, the  images are autofluorescence adjusted and then single-cell segmentation is performed using machine learning \citep{mckinley_machine_2019}.
CD8 expression data are quantified for each cell by taking the median image intensity in each cell's segmented region.
Cell values are transformed using {\tt log10(x/mean(x)+1) } \citep{harris_quantifying_2021}.
The CD8 marker values are composed of two cell populations, CD8 positive cells that are expressed and background cells that are not expressed. The cells that are CD8 positive have a higher value of CD8 than the background cells and should account for approximately 5-15\% of the cell population \citep{chen_human_2021}.
We expect that adding constraints to the CD8 positive cells component will provide a more natural fit.

\begin{figure}[!p]
\centering\includegraphics{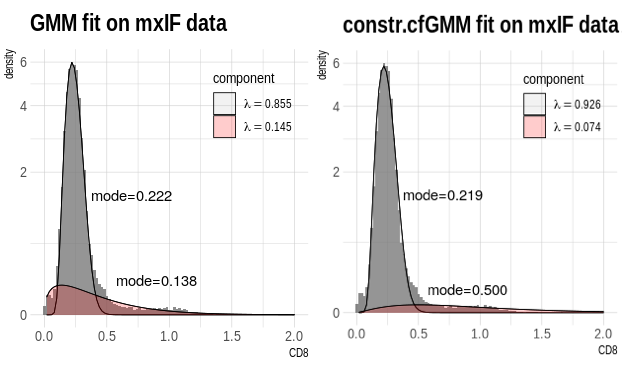}
\caption{The side-by-side comparison of GMM and cfGMM fit on mxIF CD8 data. The y-axis is on log-scale for display purposes.}
\label{fig:mxIF}
\end{figure}

The fit of GMM on the left panel of Figure \ref{fig:mxIF} shows that the two component distributions are overlapping, which contradicts our prior belief about the signal intensity of CD8$+$ cells. In contrast the constrained 2-component cfGMM has a better fit (modal constraints $[0, 0.8]$ and $[0.5, 1.5]$) We chose the constraint by looking of histogram. The fit from the constrained model, which is on the right panel of Figure \ref{fig:mxIF} is closer to our expectation, distinguishing the two components with the expressed distribution constrained to the right of the unexpressed cell.
 
The mode of the expressed group fit by constrained cfGMM lies on the boundary of constraint because the unconstrained likelihood wants to put the "expressed" component mode below the unexpressed cells.
The gamma is just an approximation to the distribution of the mIF and does not perfectly fit the data, but is useful for estimating expressed cell proportions and individual cell probabilities, $z_ik$, \eqref{eq:lambdaFormula}.

\section{Discussion}
\label{sec5}

In this paper, we developed closed-form estimator for the finite gamma mixture model and demonstrated its computational advantages compared to the existing GMM estimation method \citep{young_finite_2019}. We used the approach of \citet{ye_closed-form_2017} to derive closed-form estimators for the gamma distribution, but generalized it to $K$-component mixture models.
Having closed-form estimators made it easy to add parameter constraints on the mode of the component distributions.
Incorporating the constraints made the model converge faster than the cfGMM without constraints and the GMM.

Our goal is to use this approach to estimate cell classes in mIF data and future work will further develop the application of this model to biological problems. For example, in mIF, cell type identification is traditionally done manually, which is infeasible for a large number of slides.
There are a few methods to perform automatic cell type identification \citep{levine_data-driven_2015,geuenich_automated_2021,zhang_probabilistic_2019}, but these do not account for the skewed distribution of the data; it will be interesting to see whether the constrained cfGMM improves cell population estimation. Model fit might be improved by adding constraints on variance of the components as well. The cfGMM proposed in this paper can be a good candidate for a robust statistical method that reduces the burden of manual cell type identification.

\section{Software}
\label{sec6}

Software in the form of R code, is available on \url{https://github.com/JiangmeiRubyXiong/cfGMM}.



\section*{Acknowledgments}

{\it Conflict of Interest}: None declared.

\section*{Funding}
This research was supported by National Institutes of Health grants (R01DK103831 and U01CA215798 to K.S.L., U2CCA233291 to R.C., K.S.L., M.J.S., R01MH123563 to S.N.V.) and the Vanderbilt Ingram Cancer Center GI SPORE (P50CA236733).

\bibliographystyle{biorefs}
\bibliography{MyLibrary}

\begin{thebibliography}{99}

\bibitem[Ashburner and Friston(2005)Ashburner and
  Friston]{ashburner_unified_2005}
\textsc{Ashburner, John and Friston, Karl~J.} (2005, July).
\newblock Unified segmentation.
\newblock {\em NeuroImage\/}~\textbf{26}(3), 839--851.

\bibitem[Benaglia \emph{and others}(2009)Benaglia, Chauveau, Hunter and
  Young]{benaglia_mixtools_2009}
\textsc{Benaglia, Tatiana, Chauveau, Didier, Hunter, David and Young, Derek}.
  (2009).
\newblock mixtools: {An} {R} package for analyzing finite mixture models.
\newblock {\em Journal of statistical software\/}~\textbf{32}(6), 1--29.

\bibitem[Boos and Stefanski(2013)Boos and Stefanski]{boos_essential_2013}
\textsc{Boos, Dennis~D. and Stefanski, Leonard~A.} (2013).
\newblock {\em Essential {Statistical} {Inference}: {Theory} and {Methods}\/},
  Springer {Texts} in {Statistics}. New York: Springer-Verlag.

\bibitem[Chen \emph{and others}(2021)Chen, McKinley, Simmons, Ramirez-Solano,
  Zhu, Southard-Smith, Markham, Sheng, Drewes, Xu, Heiser, Zhou, Revetta,
  Berry, Zheng, Washington, Cai, Sears, Goldenring, Franklin, Vandekar, Roland,
  Su, Huh, Liu, Coffey, Shrubsole and Lau]{chen_human_2021}
\textsc{Chen, Bob, McKinley, Eliot~T., Simmons, Alan~J., Ramirez-Solano,
  Marisol~A., Zhu, Xiangzhu, Southard-Smith, Austin~N., Markham, Nicholas~O.,
  Sheng, Quanhu, Drewes, Julia~L., Xu, Yanwen, Heiser, Cody~N., Zhou, Yuan,
  Revetta, Frank, Berry, Lynne, Zheng, Wei, Washington, M.~Kay, Cai, Qiuyin,
  Sears, Cynthia~L., Goldenring, James~R., Franklin, Jeffrey~L., Vandekar,
  Simon, Roland, Joseph~T., Su, Timothy, Huh, Won~Jae, Liu, Qi, Coffey,
  Robert~J., Shrubsole, Martha~J.} \emph{and others}. (2021, January).
\newblock Human colorectal pre-cancer atlas identifies distinct molecular
  programs underlying two major subclasses of pre-malignant tumors.
\newblock {\em Technical Report}.
\newblock Company: Cold Spring Harbor Laboratory Distributor: Cold Spring
  Harbor Laboratory Label: Cold Spring Harbor Laboratory Section: New Results
  Type: article.

\bibitem[Dempster \emph{and others}(1977)Dempster, Laird and
  Rubin]{dempster_maximum_1977}
\textsc{Dempster, A.~P., Laird, N.~M. and Rubin, D.~B.} (1977).
\newblock Maximum {Likelihood} from {Incomplete} {Data} {Via} the {EM}
  {Algorithm}.
\newblock {\em Journal of the Royal Statistical Society: Series B
  (Methodological)\/}~\textbf{39}(1), 1--22.

\bibitem[Gerdes \emph{and others}(2013)Gerdes, Sevinsky, Sood, Adak, Bello,
  Bordwell, Can, Corwin, Dinn and Filkins]{gerdes_highly_2013}
\textsc{Gerdes, Michael~J., Sevinsky, Christopher~J., Sood, Anup, Adak,
  Sudeshna, Bello, Musodiq~O., Bordwell, Alexander, Can, Ali, Corwin, Alex,
  Dinn, Sean and Filkins, Robert~J.} (2013).
\newblock Highly multiplexed single-cell analysis of formalin-fixed,
  paraffin-embedded cancer tissue.
\newblock {\em Proceedings of the National Academy of
  Sciences\/}~\textbf{110}(29), 11982--11987.

\bibitem[Geuenich \emph{and others}(2021)Geuenich, Hou, Lee, Ayub, Jackson and
  Campbell]{geuenich_automated_2021}
\textsc{Geuenich, Michael~J., Hou, Jinyu, Lee, Sunyun, Ayub, Shanza, Jackson,
  Hartland~W. and Campbell, Kieran~R.} (2021, September).
\newblock Automated assignment of cell identity from single-cell multiplexed
  imaging and proteomic data.
\newblock {\em Cell Systems\/}.

\bibitem[Harris \emph{and others}(2021)Harris, McKinley, Roland, Liu,
  Shrubsole, Lau, Coffey, Wrobel and Vandekar]{harris_quantifying_2021}
\textsc{Harris, C.~R., McKinley, E.~T., Roland, J.~T., Liu, Q., Shrubsole,
  M.~J., Lau, K.~S., Coffey, R.~J., Wrobel, J. and Vandekar, S.~N.} (2021,
  July).
\newblock Quantifying and correcting slide-to-slide variation in multiplexed
  immunofluorescence images.
\newblock {\em bioRxiv\/}, 2021.07.16.452359.

\bibitem[Karlis and Xekalaki(2003)Karlis and Xekalaki]{karlis_choosing_2003}
\textsc{Karlis, Dimitris and Xekalaki, Evdokia}. (2003).
\newblock Choosing initial values for the em algorithm for finite mixtures.
\newblock {\em Computational Statistics \& Data Analysis\/}~\textbf{41}(3-4),
  577--590.

\bibitem[Khalili \emph{and others}(2007)Khalili, Potter, Yan, Li, Gray, Huang
  and Lin]{khalili_gamma-normal-gamma_2007}
\textsc{Khalili, Abbas, Potter, Dustin, Yan, Pearlly, Li, Lang, Gray, Joe,
  Huang, Tim and Lin, Shili}. (2007).
\newblock Gamma-normal-gamma mixture model for detecting differentially
  methylated loci in three breast cancer cell lines.
\newblock {\em Cancer Informatics\/}~\textbf{3}, 117693510700300012.

\bibitem[Lakshmi and Vaidyanathan(2016)Lakshmi and
  Vaidyanathan]{lakshmi_parameter_2016}
\textsc{Lakshmi, R~Vani and Vaidyanathan, VS}. (2016).
\newblock Parameter estimation in gamma mixture model using normal-based
  approximation.
\newblock {\em J. Stat. Theory Appl.\/}~\textbf{15}(1), 25--35.

\bibitem[Levine \emph{and others}(2015)Levine, Simonds, Bendall, Davis, El-ad,
  Tadmor, Litvin, Fienberg, Jager and Zunder]{levine_data-driven_2015}
\textsc{Levine, Jacob~H., Simonds, Erin~F., Bendall, Sean~C., Davis, Kara~L.,
  El-ad, D.~Amir, Tadmor, Michelle~D., Litvin, Oren, Fienberg, Harris~G.,
  Jager, Astraea and Zunder, Eli~R.} (2015).
\newblock Data-driven phenotypic dissection of {AML} reveals progenitor-like
  cells that correlate with prognosis.
\newblock {\em Cell\/}~\textbf{162}(1), 184--197.

\bibitem[McKinley \emph{and others}(2019)McKinley, Roland, Franklin, Macedonia,
  Vega, Shin, Coffey and Lau]{mckinley_machine_2019}
\textsc{McKinley, Eliot~T., Roland, Joseph~T., Franklin, Jeffrey~L., Macedonia,
  Mary~Catherine, Vega, Paige~N., Shin, Susie, Coffey, Robert~J. and Lau,
  Ken~S.} (2019, October).
\newblock Machine and deep learning single-cell segmentation and quantification
  of multi-dimensional tissue images.
\newblock {\em bioRxiv\/}, 790162.

\bibitem[McKinley \emph{and others}(2017)McKinley, Sui, Al-Kofahi, Millis,
  Tyska, Roland, Santamaria-Pang, Ohland, Jobin, Franklin, Lau, Gerdes and
  Coffey]{mckinley_optimized_2017}
\textsc{McKinley, Eliot~T., Sui, Yunxia, Al-Kofahi, Yousef, Millis, Bryan~A.,
  Tyska, Matthew~J., Roland, Joseph~T., Santamaria-Pang, Alberto, Ohland,
  Christina~L., Jobin, Christian, Franklin, Jeffrey~L., Lau, Ken~S., Gerdes,
  Michael~J.} \emph{and others}. (2017, June).
\newblock Optimized multiplex immunofluorescence single-cell analysis reveals
  tuft cell heterogeneity.
\newblock {\em JCI Insight\/}~\textbf{2}(11).

\bibitem[Solomon \emph{and others}(2006)Solomon, Butman and
  Sood]{solomon_segmentation_2006}
\textsc{Solomon, Jeffrey, Butman, John~A. and Sood, Arun}. (2006, December).
\newblock Segmentation of brain tumors in {4D} {MR} images using the hidden
  {Markov} model.
\newblock {\em Computer Methods and Programs in Biomedicine\/}~\textbf{84}(2),
  76--85.

\bibitem[Stacy(1962)Stacy]{stacy_generalization_1962}
\textsc{Stacy, E.~W.} (1962).
\newblock A {Generalization} of the {Gamma} {Distribution}.
\newblock {\em The Annals of Mathematical Statistics\/}~\textbf{33}(3),
  1187--1192.
\newblock Publisher: Institute of Mathematical Statistics.

\bibitem[Ye and Chen(2017)Ye and Chen]{ye_closed-form_2017}
\textsc{Ye, Zhi-Sheng and Chen, Nan}. (2017).
\newblock Closed-form estimators for the gamma distribution derived from
  likelihood equations.
\newblock {\em The American Statistician\/}~\textbf{71}(2), 177--181.

\bibitem[Young \emph{and others}(2019)Young, Chen, Hewage and
  Nilo-Poyanco]{young_finite_2019}
\textsc{Young, Derek~S., Chen, Xi, Hewage, Dilrukshi~C. and Nilo-Poyanco,
  Ricardo}. (2019, December).
\newblock Finite mixture-of-gamma distributions: estimation, inference, and
  model-based clustering.
\newblock {\em Advances in Data Analysis and Classification\/}~\textbf{13}(4),
  1053--1082.

\bibitem[Zhang \emph{and others}(2019)Zhang, O'Flanagan, Chavez, Lim, Ceglia,
  McPherson, Wiens, Walters, Chan, Hewitson, Lai, Mottok, Sarkozy, Chong, Aoki,
  Wang, Weng, McAlpine, Aparicio, Steidl, Campbell and
  Shah]{zhang_probabilistic_2019}
\textsc{Zhang, Allen~W., O'Flanagan, Ciara, Chavez, Elizabeth~A., Lim, Jamie
  L.~P., Ceglia, Nicholas, McPherson, Andrew, Wiens, Matt, Walters, Pascale,
  Chan, Tim, Hewitson, Brittany, Lai, Daniel, Mottok, Anja, Sarkozy,
  Clementine, Chong, Lauren, Aoki, Tomohiro, Wang, Xuehai, Weng, Andrew~P.,
  McAlpine, Jessica~N., Aparicio, Samuel, Steidl, Christian, Campbell,
  Kieran~R.} \emph{and others}. (2019, October).
\newblock Probabilistic cell-type assignment of single-cell {RNA}-seq for tumor
  microenvironment profiling.
\newblock {\em Nature Methods\/}~\textbf{16}(10), 1007--1015.

\bibitem[Zhang \emph{and others}(2001)Zhang, Brady and
  Smith]{zhang_segmentation_2001}
\textsc{Zhang, Y., Brady, M. and Smith, S.} (2001, January).
\newblock Segmentation of brain {MR} images through a hidden {Markov} random
  field model and the expectation-maximization algorithm.
\newblock {\em IEEE Transactions on Medical Imaging\/}~\textbf{20}(1), 45--57.

\end{thebibliography}

\begin{table}[!p]
\tblcaption{The proportion of convergence in results among 1000 simulations. constr.$=$constrained, cfGMM$=$closed-form gamma mixture model.
\label{tab:convergence}}
{\tabcolsep=4.25pt
\begin{tabular}{l c c c|c c c } 
\toprule
& \multicolumn{6}{c}{\textbf{Convergence Proportion}} \\
\cmidrule(l){2-7}
\textbf{Sample Size}&\multicolumn{3}{c|}{\textbf{2 Component}} & \multicolumn{3}{c }{\textbf{3 Component}} \\
  &constr.cfGMM & cfGMM &  GMM & constr.cfGMM & cfGMM &  GMM\\
 \midrule
 100&1 & 1 & 0.992 & 1 & 1 & 0.608\\ 
 1000&1 & 1 & 0.999 & 1 & 1 & 0.706\\ 
 10000&1 & 1 & 0.998 & 1 & 1 & 0.875\\  
 \bottomrule
\end{tabular}}
\end{table}

\end{document}